\documentclass[12pt]{article}
\pdfoutput=1
\usepackage[paperwidth=8.5in,paperheight=11.in,
body={6.5in,9.5in},top=1.in,bottom=1.in,
left=1.in,right=1.in]{geometry}

\usepackage[bibstyle=nature,citestyle=numeric-comp,sorting=none]{biblatex}
\usepackage{graphicx}
\usepackage{xspace}
\usepackage{amssymb}
\usepackage{amsmath}

\newcommand{\etal}{\textit{et al.}}
\newcommand{\ie}{i.e.\ }
\newcommand{\be}{\begin{equation}}
\newcommand{\ee}{\end{equation}}

\newcommand{\fig}[1]{fig.~\ref{#1}}
\newcommand{\Fig}[1]{Fig.~\ref{#1}}

\newcommand{\ohm}{$\Omega$\xspace}
\newcommand{\micron}{$\mu$m\xspace}

\begin{document}
\setlength{\baselineskip}{24pt}

\newcommand{\jpl}{${}^*$}
\newcommand{\cit}{${}^\dagger$}
\newcommand{\titlefont}[1]{{\Large{\textbf{\textsf{#1}}}}}
\newcommand{\auth}[1]{{\textbf{\textsf{#1}}}}
\newcommand{\addr}[1]{{\small \textit{#1}}}

\begin{center}
\titlefont{A Wideband, Low-Noise Superconducting Amplifier with High Dynamic Range}\\
\vspace{0.1in}
\auth{Byeong Ho Eom}\cit,
\auth{Peter K. Day}\jpl,
\auth{Henry G. LeDuc}\jpl,
\auth{Jonas Zmuidzinas}\cit\jpl,\\
\vspace{0.1in}
{\small (Dated: \today)}
\end{center}

\vspace{0.1in}

\noindent
{\cit}\addr{California Institute of Technology,
Pasadena CA 91125, USA}

\noindent
{\jpl}\addr{Jet Propulsion Laboratory, California Institute of Technology, Pasadena CA 91109, USA}

\vspace{0.2in}

\noindent \textbf{ %
Amplifiers are ubiquitous in electronics and play a fundamental
role in a wide range of scientific measurements.
From a user's perspective, an ideal amplifier has very low noise,
operates over a broad frequency range, and has a high dynamic range -
it is capable of handling strong signals with little distortion.
Unfortunately, it is difficult
to obtain all of these characteristics simultaneously.
For example,
modern transistor amplifiers offer multi-octave bandwidths and excellent dynamic range.
However, their noise remains far above the fundamental limit set by the
uncertainty principle of  quantum mechanics.\cite{Caves_1982}
Parametric amplifiers, which predate transistor amplifiers and are  widely used in optics,
exploit a nonlinear response to transfer power from a strong pump tone
to a weak signal.  If the nonlinearity is purely reactive, \ie nondissipative,
in theory the amplifier noise can reach the quantum-mechanical limit.\cite{Louisell_1961}
Indeed, microwave frequency superconducting Josephson
parametric amplifiers\cite{
Zimmer_1967,
Moshovich_1990}
do approach the quantum limit,
but generally are narrow band and have very limited dynamic range.
In this paper, we describe a superconducting parametric amplifier that overcomes
these limitations.
The amplifier is very simple, consisting only of a patterned metal film on a dielectric substrate,
and relies on the nonlinear kinetic inductance of
a superconducting transmission line.
We measure gain extending
over 2~GHz on either side of an 11.56~GHz pump tone, and
we place an upper limit on the added noise of the amplifier of 3.4 photons at 9.4~GHz.
Furthermore, the dynamic range is very large, comparable to
microwave transistor amplifiers, and the concept can be
applied throughout the microwave, millimeter-wave
and submillimeter-wave bands.
}

%
Over the past decade, the combination of high-performance superconducting microresonators
and low-noise, microwave frequency cryogenic transistor amplifier readouts has proven to be particularly powerful for a wide range of applications including photon detection
and quantum information experiments.\cite{Day_2003,Wallraff_2004,Zmuidzinas_2012}
These developments have generated strong renewed interest in superconducting amplifiers that achieve even lower readout noise.
\cite{Castellanos_2008, Yamamoto_2008, Bergeal_2010, Spietz_2010,
Hover_2011}.
Most of these devices are parametric amplifiers that
make use of the nonlinear inductance of the Josephson junction,
which is almost ideally reactive with little dissipation below the  critical current $I_c$.
As a result, Josephson paramps
can be exquisitely sensitive, approaching
the standard quantum limit of half a photon $\hbar \omega /2$ of added noise
power per unit bandwidth in the standard case
when both quadratures of a signal at frequency $\omega$ are amplified equally.
Here $\hbar$ is Planck's constant divided by $2 \pi$.
Even less noise is possible in situations when only one quadrature is amplified.\cite{Caves_1982}.
In comparison, the added noise of cryogenic transistor amplifiers is typically
10-20 times the quantum limit.\cite{Pospieszalski_2005}
However, the dynamic range of Josephson paramps is regulated
by the Josephson energy $E_J = \hbar I_c / 4 \pi e$
(here $e$ is the electron charge) to values that are far lower than for transistor amplifiers.
Furthermore, as in optical parametric oscillators in which
light passes many times through a nonlinear medium,
previous superconducting paramps generally use resonant circuits to
enhance the effective nonlinearity in order to achieve high gain.
Consequently, amplification is achieved over a narrow instantaneous
frequency range,  typically of order a few MHz, vs. $\sim 10\ \mathrm{GHz}$
for transistor amplifiers.
This results in a relatively slow response time that
can hinder observation of time-dependent phenomena, e.g. quantum jumps.\cite{Vijay_2011}
Also, as with superconducting quantum interference devices (SQUIDs),
the combination of a limited dynamic range and a limited bandwidth
results in a low Shannon information capacity and
limits the utility of Josephson paramps for multiplexed readout of detector
arrays.\cite{Irwin_2009}

%
Instead of using a resonator, the optical or electrical path may be unfolded into a long nonlinear transmission line or waveguide.
This results in a traveling wave paramp\cite{Cullen_1958, Tien_1958} which has a very broad
intrinsic bandwidth.
However, a low-dissipation medium that is sufficiently nonlinear over a realizable length must be found. %
At visible and infrared wavelengths, these requirements are met in optical fibers\cite{Hansryd02} and silicon waveguides\cite{Foster_2006}  through the nonlinear
process of four-wave mixing (FWM) that results from the intensity dependence of the refractive index, \ie the Kerr effect.
Fiber paramps achieve high ($> 60$~dB)
gain\cite{Hansryd02} and single-quadrature versions have exhibited noise levels
below the standard quantum limit.\cite{Tong_2011}
An analogous microwave device using a metamaterial of numerous Josephson junctions
embedded in a transmission line\cite{Sweeny85,Yurke96} has been proposed
and investigated, but this design has not yet resulted in a practical amplifier.

%
Like a Josephson junction, a thin superconducting wire behaves as a nondissipative
inductor for currents below a critical current $I_c$.
The critical current is therefore an obvious
scale for nonlinear behavior in both junctions and wires,
although $I_c$ for a wire is typically
orders of magnitude larger than for a junction.
Indeed, the phenomenological Ginzburg-Landau theory
and the microscopic BCS theory\cite{Parmenter62, Anthore_2003}
both predict the nonlinearity of the kinetic inductance of superconductors.
In practice this nonlinearity is usually weak, though
resonant paramps based on this effect
have been proposed and investigated\cite{Landauer_patent_1963,Tholen_2007}.
Here we show that
the use of a high-resistivity superconductor such as TiN\cite{Leduc_2010}
or NbTiN results in a kinetic inductance nonlinearity that
is sufficient to allow parametric gain in a practical,
realizable traveling wave geometry.

%
The current variation of the kinetic inductance of a superconducting wire
is expected to be quadratic to lowest order,
\ie $L_k(I) \approx L_k(0)\left[ 1 + (I/I_\ast)^2 \right]$,
just as in the case of Josephson junctions.
The Mattis Bardeen theory gives
$L_k(0) = \hbar R_n  / \pi \Delta$
for a wire with normal state resistance $R_n$ and superconducting
gap parameter $\Delta$ and
whose transverse dimensions are small enough so that
the current distribution is approximately uniform.
$I_\ast$ is comparable to $I_c$,
and can be roughly estimated by equating the kinetic energy of the Cooper pairs $L_k I^2/2$
to the pairing energy
$E_p = 2 N_0 \Delta^2 V$, where $N_0$ is the density of states at the Fermi level and $V$ is the volume.
The resulting expression for $I_\ast ^2$ is proportional to $1 / R_n$.
The phase velocity of the transmission line is
$v_{ph} = 1 / \sqrt{\mathcal{L}\mathcal{C}} \approx v_{ph}(0)(1 - \alpha I^2 / 2I_\ast^2)$, where $\mathcal{L}$ and $\mathcal{C}$ are the total inductance
and capacitance per unit length, and $\alpha$ is the ratio of kinetic inductance to total inductance.  The magnitude of the Kerr effect in the line is therefore
proportional to $\alpha / I_\ast^2$ and is enhanced in films with high normal state resistivity $\rho_n$
due both to large $\alpha$ and small $I_\ast$.
The TiN and NbTiN films produced in our laboratory have
$\rho_n \approx 100\ \mu\Omega\,\mathrm{cm}$,
nearly three orders of magnitude larger than for typical aluminum films,
and have very low microwave loss in the superconducting state.\cite{Leduc_2010}
The high resistivity also results in a large penetration depth,\cite{Zmuidzinas_2012}
in the range $2-20\,\mu\mathrm{m}$ depending on thickness (20-50~nm typically) and
critical temperature $T_c$,
so a uniform current density is readily achieved in our
micron-scale wires patterned by optical lithography.

\begin{figure}
A\hspace{1.7in}C\\
 \includegraphics[width=1.7in]{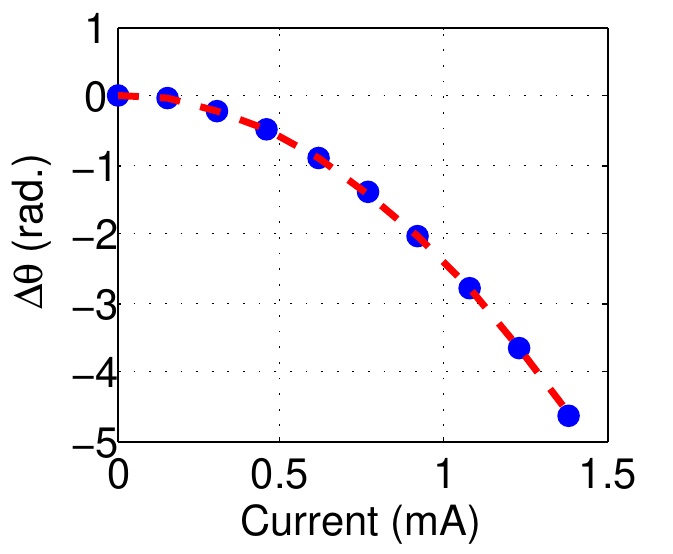}
 \includegraphics[width=2.1in]{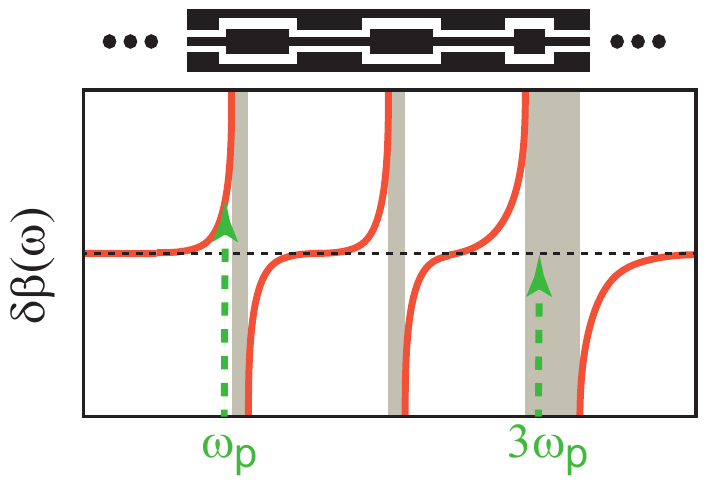}
\\
B\\
 \includegraphics[width=3.7in]{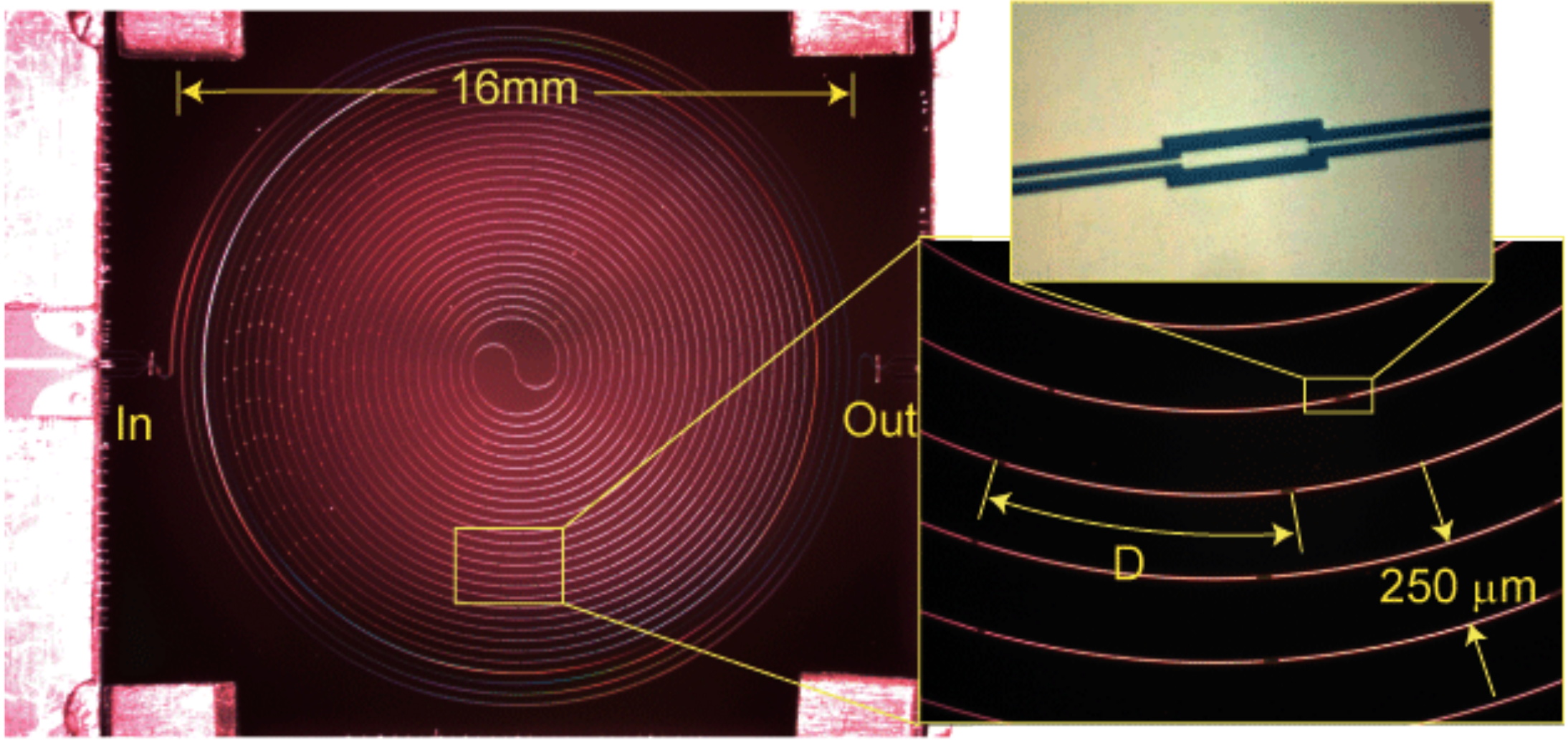}
\caption{Eom \etal}
\label{fig:device}
\end{figure}

%
The nonlinear kinetic inductance is illustrated in \fig{fig:device}, which shows the phase shift of a
microwave tone passing through an NbTiN transmission line
as a function of the DC current flowing through it.  In fact,
such a current-controlled phase shifter has been previously proposed\cite{Anlage_89}
but not successfully demonstrated due to an increase in the microwave dissipation with DC current.
Similar dissipative behavior has also been noted
in superconducting thin film
resonators (see supplementary information).
In contrast,
no increase in dissipation was observed for the range of phase shifts displayed in \fig{fig:device}.
Also, provided that the temperature is kept well below the critical temperature $T_c$,
our TiN\cite{Leduc_2010} and NbTiN microresonators remain nondissipative for microwave currents sufficiently strong to develop a significant reactive nonlinearity.
These results are described in the supplementary information.

The parametric gain produced through FWM can be calculated using coupled mode
equations that have been developed to describe optical fiber
paramps (see supplementary information).
In general, FWM may
involve two separate pump tones, but we consider only the degenerate case where the pump
frequencies are equal. The  equations then
describe the interaction of the pump at angular frequency $\omega_p = 2 \pi f_p$,
signal at $\omega_s$, and the generated idler tone at $\omega_i = 2\omega_p - \omega_s$.
For a dispersionless
line, which is a good approximation for a uniform superconducting transmission line
at frequencies well below the gap frequency $2 \Delta / h$, the
signal and idler gains for $f_s \approx f_p$
are $G_s = 1 + (\Delta \theta)^2$ and $G_i = (\Delta \theta)^2$, where
$\Delta \theta$ is the nonlinear phase shift, in radians, of the pump tone due to its own AC current.
\Fig{fig:device} shows that phase shifts of several radians can be achieved in response to a DC current.
Comparable phase shifts can also be achieved in response to RF currents,
as described in the supplementary information.
Dispersion, due either to material or waveguide  properties,
controls the phase slippage between the waves as they propagate,
which in turn determines whether the signal is amplified or deamplified.
The linear dispersion
$\Delta \beta = \beta (\omega_S) +  \beta (\omega_I) - 2 \beta(\omega_P)$
involves a difference of the propagation constants $\beta(\omega)$ for small-amplitude waves
at the signal, idler and pump frequencies, and vanishes for a dispersionless line
obeying $\beta(\omega) = \omega / \bar{c}$.
The coupled-mode equations predict that maximum gain occurs
when $\Delta \beta = -2 \Delta \theta / L$ because this value of linear dispersion
compensates for the phase slippage that arises from the nonlinearity.
Furthermore, the resulting gain $G_s = \exp{(2\Delta\theta)}/4$ varies exponentially with line length $L$ rather than quadratically\cite{Stolen_1982, Hansryd02}.  For much larger or smaller values of $\Delta\beta$, loss of phase match yields low gain.

%
In fact, some amount of dispersion is necessary, because a dispersionless Kerr medium leads to generation of a shock front when $\Delta \theta \gtrsim 1$,
preventing significant parametric gain.\cite{Landauer_1960}
Indeed, a superconducting line generates the third harmonic $3 f_p$
due to the voltage term $I^2 dI/dt$ arising from the nonlinear inductance.
This is the first step in the formation of the shock front: once the $3 f_p$
harmonic is present, other harmonics can be generated.
This problem is dealt with in a simple manner
using dispersion engineering, leading to the device shown in \fig{fig:device}.
Periodic perturbations are included in the coplanar waveguide (CPW)
transmission line with a separation corresponding to half of a wavelength
at frequency $f_\mathrm{per}$, with $f_\mathrm{per}$ slightly larger than $3 f_p$.
Much like an electronic or photonic bandgap,
this results in a stop band centered at $f_\mathrm{per}$
that includes $3 f_p$, blocking harmonic generation.
We also slightly alter every third perturbation, resulting in weak stop bands
around $f_\mathrm{per}/3$ and  $2 f_\mathrm{per}/3$ (\fig{fig:device}c).
This gives rise to localized dispersion features near these frequencies;
fine-tuning of the pump frequency $f_p$ in the vicinity of these features
allows the optimum value of $\Delta \beta$ to be achieved for a wide range of signal frequencies.
For the device shown in \fig{fig:device}, the stop bands occur at at multiples of
$f_\mathrm{per}/3 \approx \,$5.9~GHz,  not far from the design value of 5~GHz.
We chose to operate the the device with the pump tuned near
$2 f_\mathrm{per}/3 \approx 11.8\,$GHz because this
dispersion feature was stronger; note that a strong stop band also occurs at $3\times 11.8\,$GHz,
which prevents formation of pump harmonics.

\begin{figure}
 \includegraphics[width=3.8in]{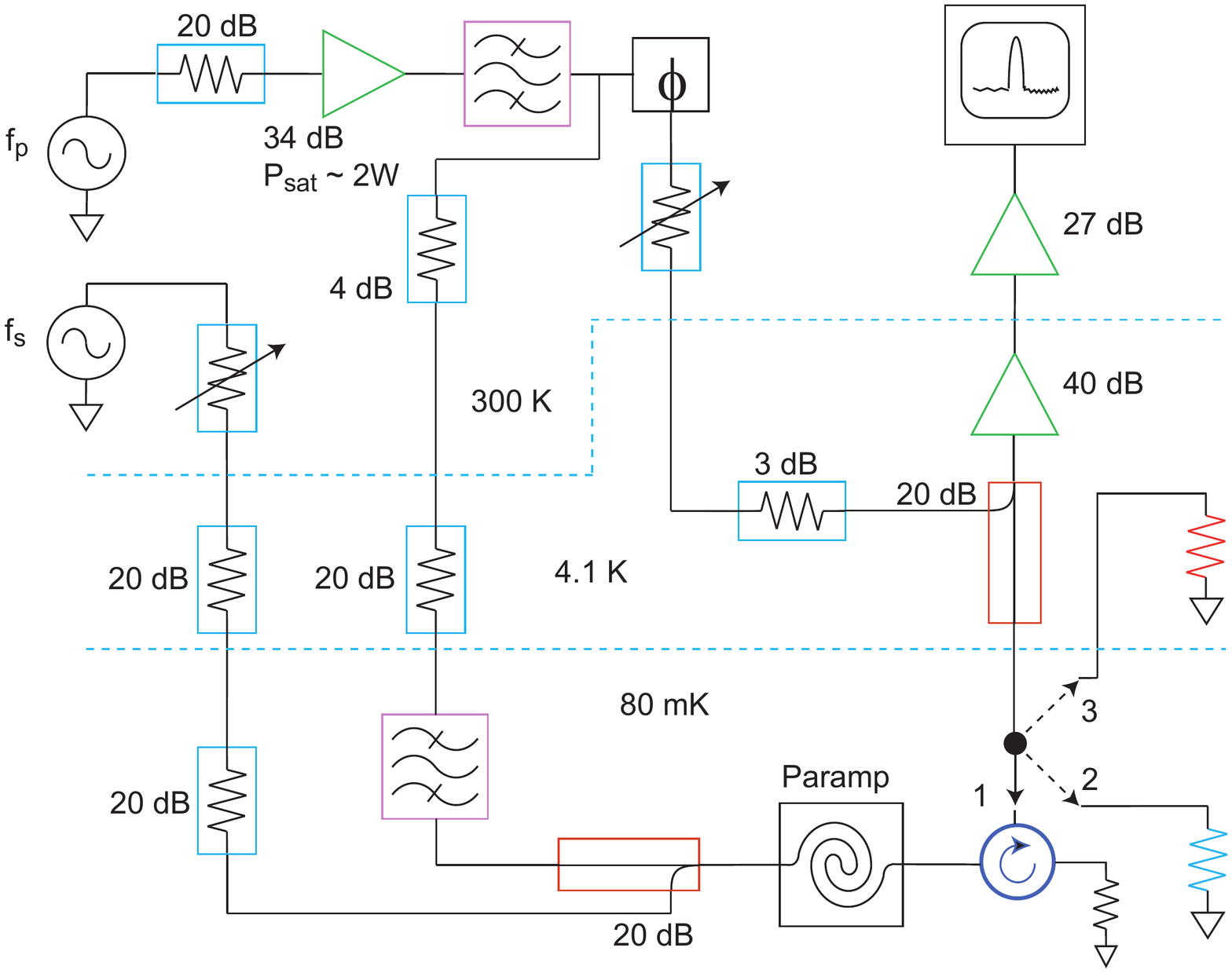}
\caption{Eom \etal}
\label{fig:circuit}
\end{figure}

\begin{figure}
 \includegraphics[width=3.5in]{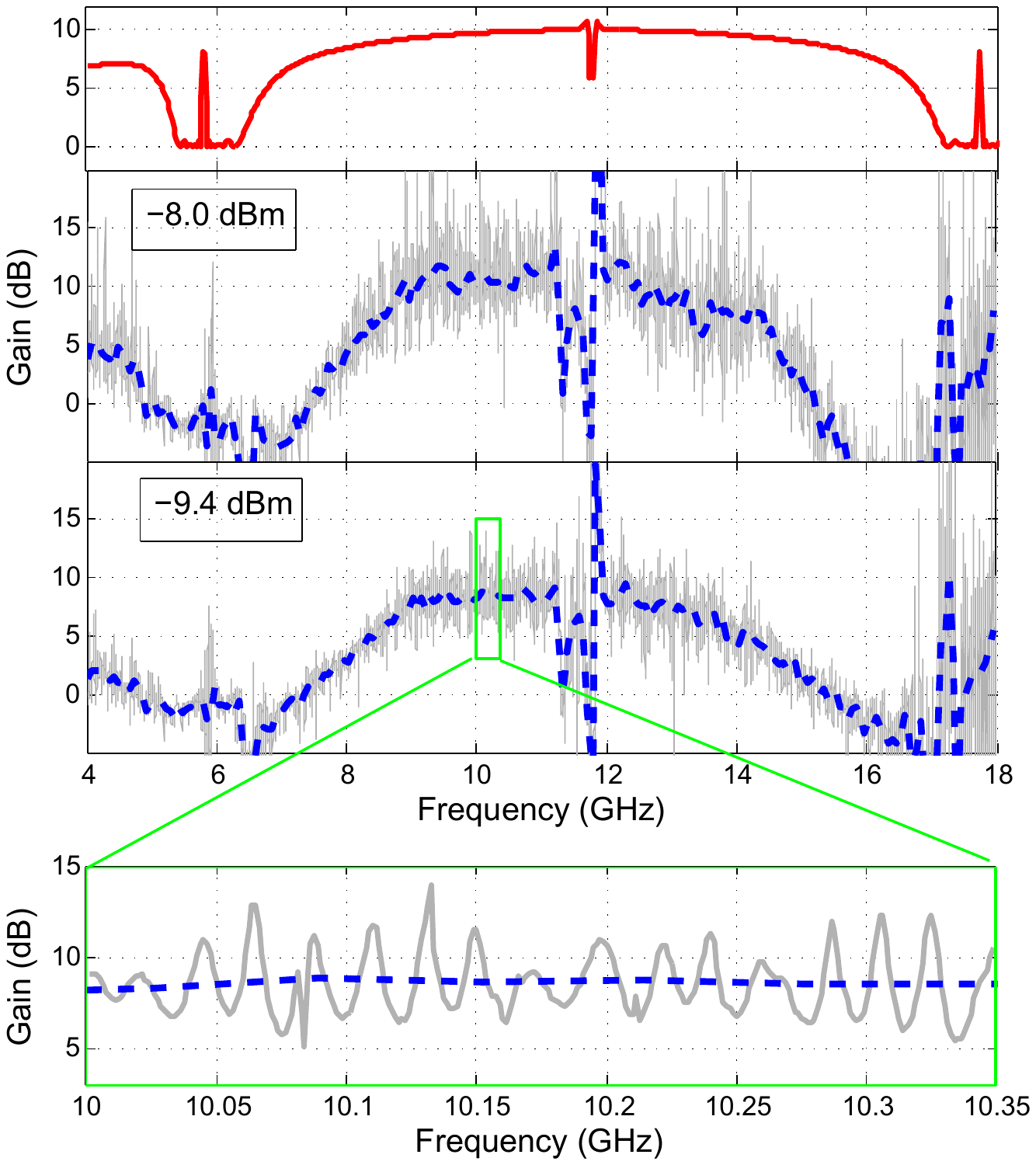}
\caption{Eom \etal}
\label{fig:gain}
\end{figure}

%
We measured the gain of the amplifier
with the pump tuned to 11.56~GHz
using the circuit shown in \fig{fig:circuit}, excluding the bandpass filter.
The gain increases with pump power until a critical pump power is reached,
which we identify with the onset of nonlinear dissipation in the line.
Significant gain is observed over a frequency range of approximately 8~GHz to 14~GHz, with a notch around the pump frequency (\fig{fig:gain}).
A theoretical gain curve (\fig{fig:gain}, top), generated by integrating the coupled mode equations and including a model of the loaded line dispersion (see supplementary info), is in rough agreement with the average gain.
However, there is a fine-scale variation of the gain that is approximately periodic with signal frequency,
and this gain ripple increases with pump power and has a divergent
behavior around the critical pump power.
The average gain at the higher pump power shown is 10~dB,
with several peaks extending above 20~dB.
From the frequency  spacing, it is clear that the gain ripple is due to standing waves
created by reflections at the ends of the line. The reflections are probably
caused by the non-optimal on-chip tapered impedance transformers,
or possibly the wirebond transitions or other components near the device.

\begin{figure}
 \includegraphics[height=2.4in]{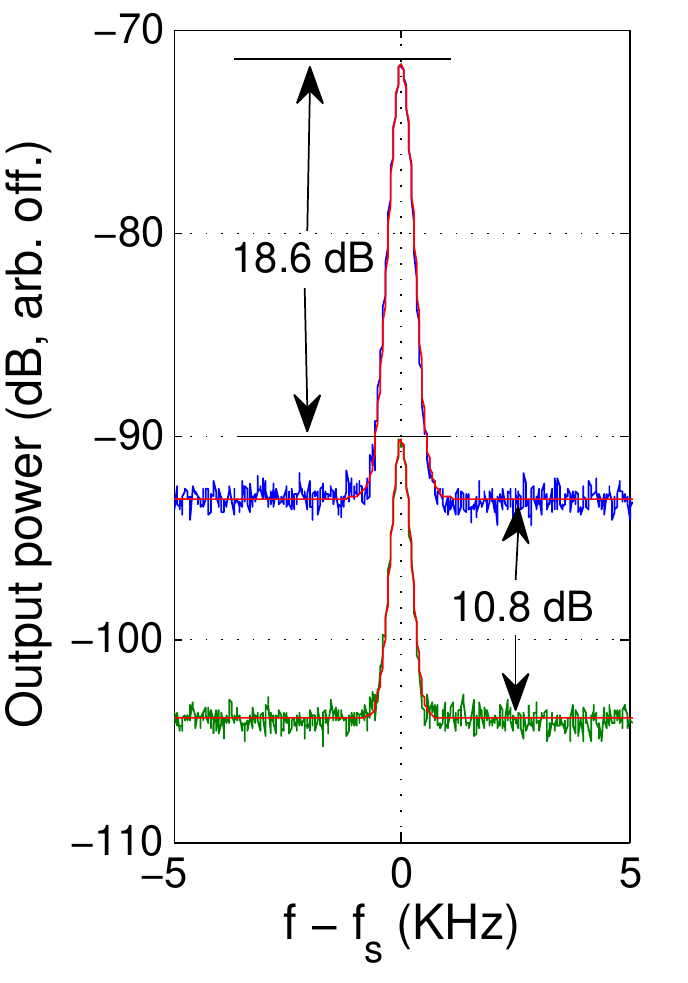}
 \includegraphics[height=2.4in]{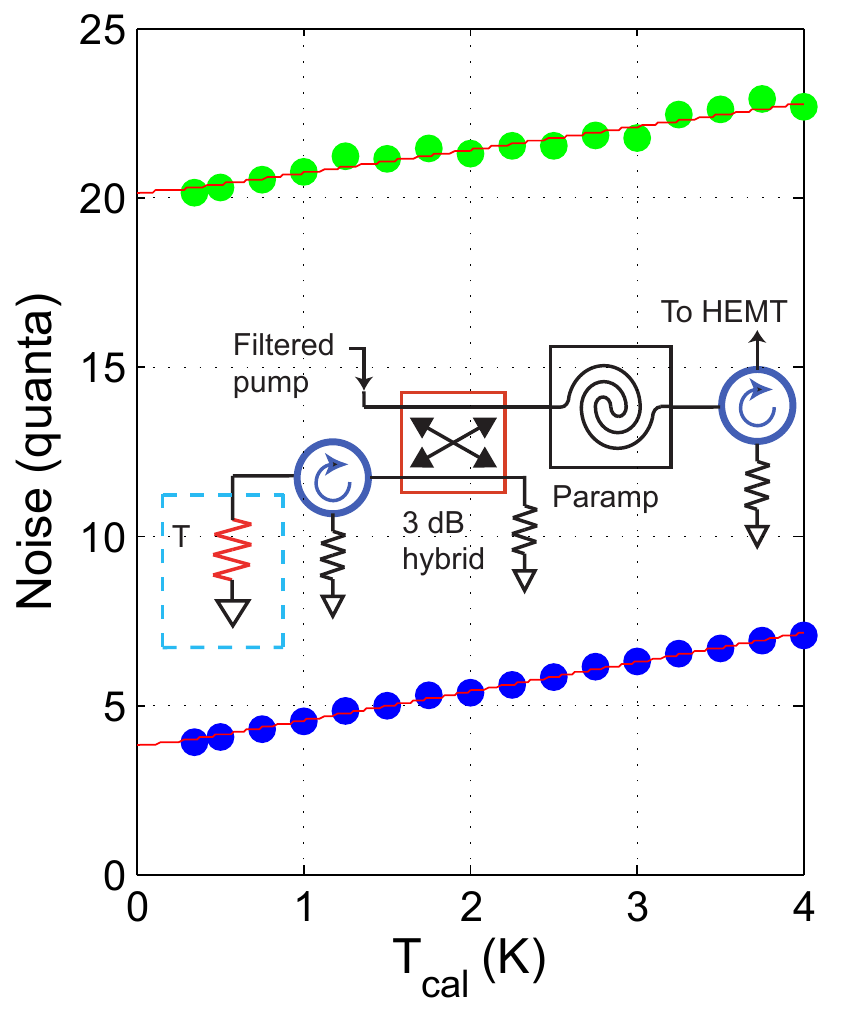}
\caption{Eom \etal}
\label{fig:SNR}
\end{figure}

%
For the purpose of measuring the noise of the paramp, we chose a signal frequency of
$f_s = 9.37$~GHz,
close to a gain peak .
The output power of the amplifier chain was measured
in a narrow band around $f_s$ using a spectrum analyzer.
A cryogenic switch  (\fig{fig:circuit}) was used to calibrate the system.
The added noise power per unit bandwidth of the high electron mobility transistor (HEMT) at 4.1~K
was measured to be $A_\mathrm{HEMT} = 16.1$, in photon units  ($\hbar \omega$),
by switching between between 50~\ohm resistors on the 4.1~K and base temperature stages of the
dilution refrigerator (switch positions 2 and 3).
With the paramp connected to the HEMT (position 1),
we make measurements with the pump tone on and off (\fig{fig:SNR}).
Because the gain peak shifts to lower frequency when the pump is applied,
we slightly reduce the signal frequency to remain at the peak.
Since the transmission of the measurement system should remain constant
over this small frequency interval, the increase in the output signal power gives the gain of the paramp, $G_\mathrm{PA} =  18.6$~dB.
It is evident that the paramp is considerably less noisy than the HEMT,
because the signal-to-noise ratio improves by 7.8~dB with the pump turned on (\fig{fig:SNR}).
The paramp noise may be quantitatively determined from
the ratio
$R = (n_\mathrm{on} - n_0) / (n_\mathrm{off} - n_0)$, where
$n_\mathrm{on}$ and $n_\mathrm{off}$ are the measured
noise levels with pump on and off, and $n_0$ is the noise from sources after the HEMT
and is found by turning off the HEMT.
However,
with the paramp connected to the HEMT (switch position 1) and no pump applied,
we find that the noise floor is slightly higher than with the HEMT connected to the cold load.
This increase in the system added noise is $A_\mathrm{sys} = 2.9$.
We have not yet determined the origin of this noise and therefore we do not know
the gain $G_\mathrm{sys}$ this noise experiences when the pump is turned on.
As a result, when we calculate the noise added by the paramp,
\be
A_\mathrm{PA} = \frac{R-1}{G'_\mathrm{PA}} A_\mathrm{H}
+ \frac{R - G_\mathrm{sys}}{G'_\mathrm{PA}} A_\mathrm{sys}
+ \frac{R}{2 G'_\mathrm{PA}} - \frac{1}{2}
\ee
we obtain a range of values $1.1 \le A_\mathrm{PA} \le 3.4$,
corresponding to $1 \le G_\mathrm{sys} \le G'_\mathrm{PA}$.
Here $G'_\mathrm{PA}$ accounts for the isolator loss (\fig{fig:circuit}).

%
To confirm the noise measurement, we used a variable-temperature 50~$\Omega$ resistor connected to the paramp input through a 3~dB hybrid coupler (\fig{fig:SNR}).
The slope of the linear relationship between output noise and resistor temperature calibrates
the gain of the system, and extrapolating this relationship to zero resistor temperature
gives the added noise of the amplifier plus any additional input noise.  If we assume
that only vacuum noise $\hbar \omega /2$ is present at the input, the added noise of the paramp is
$3.3 \pm 0.2\,$ photons, in agreement with the previous measurement.

%
The dynamic range of the paramp was investigated when operating
with 18~dB signal gain under the same conditions as used for the noise measurement.
The system gain compressed by 1~dB for a signal power at the paramp input of around 100~pW.
However, measurements made with the pump turned off show that the HEMT, rather than the paramp, is reponsible for this gain compression.
Theoretically, saturation of the paramp occurs when the amplified signal power becomes a significant fraction of the pump power, around 100~$\mu$W for this device, so that the pump becomes depleted.

%
In summary, we have demonstrated a simple and robust superconducting
amplifier with very low noise, wide bandwidth, and high dynamic range.
In fact, given the extremely low dissipation of the superconductors used,\cite{Leduc_2010}
it seems likely that the amplifier is operating very near the quantum limit and that
the added noise is due to imperfections of the present measurement system.
For example, the image frequency $f_I = 13.75$~GHz is outside the bandwidth of our isolator, so noise
from the HEMT may leak back toward the paramp at that frequency and contribute to noise at $f_s$.
Straightforward design improvements should allow high ($> 20$~dB) gain to be achieved over
an octave of instantaneous bandwidth.
Periodic loading structures can readily be designed for operating frequencies in the microwave,
millimeter-wave, and submillimeter-wave bands, potentially approaching the gap frequency of the
superconducting film ($2\Delta / h \approx 1.4\,$THz for NbTiN).
The lower frequency limit is determined only by the length of transmission line that can be fabricated.
 By applying a DC current bias, the amplifier can also be operated in a three-wave mixing mode, where the pump is at twice the  average of the signal and idler frequencies.
More generally, we hope that our demonstration will serve as a clear illustration of the remarkable
nonlinear properties of highly resistive superconductors and will stimulate development of
a much broader set of applications, just as was the case for nonlinear optics.

\printbibliography

\vspace{0.2in} \noindent \textbf{Acknowledgements}

\noindent The research was carried out at the Jet Propulsion Laboratory, California Institute of Technology, under a contract with the National Aeronautics and Space Administration and has been supported in part by NASA (Science Mission
directorate), the Keck Institute for Space Studies and the JPL Research and
Technology Development program.

\bigskip

\noindent Correspondence and requests for materials should be
addressed to P.D. \newline (e-mail: Peter.K.Day@jpl.nasa.gov).
\newpage

\noindent \textbf{Figure Legends}

\noindent\textbf{Figure \ref{fig:device}  Phase response to DC current
and amplifier design.  a}, This plot illustrates the nonlinearity of the kinetic inductance.
A NbTiN coplanar waveguide (CPW) transmission line (shown in panel \textbf{b}) was measured in transmission
using a microwave network analyzer. The total phase length was 670~radians at 4~GHz. Using bias tees,
a DC current was passed down the center conductor.  The resulting microwave
phase shift (measured at 4~GHz) displays a quadratic dependence with current.
No comparable effect occurred when adjusting the voltage of the center
strip relative to the ground planes.
This shows that the kinetic inductance has a nonlinear behavior
that is well described by $\delta L_{kin} \propto I^2$.
\textbf{b}, A picture of the amplifier (left) which consists of
a 0.8~m length of NbTiN CPW line arranged in a double spiral to reduce resonances due to
coupling between adjacent lines. The thickness of the line is 35~nm and the center
conductor and gap widths are 1~\micron.  At the input and output of the line, the CPW geometry tapers
from center strip and gap widths of 30~\micron and 5~\micron to adiabatically transform the
characteristic impedance of the line from close to 50~\ohm to 300~\ohm.  The line
is periodically loaded by widening a short section after every length $D = 877$\micron as shown on
the right, producing the stop band and dispersion characteristics.  The phase velocity on the
line is 0.1~c due to its large kinetic inductance.
\textbf{c}, An illustration of the
the effect of the periodic loading pattern (shown schematically) on the transmission of an infinite transmission
line.  The gray regions represent stop bands;  waves in these frequency ranges decay evanescently.
The graph represents the difference between the propagation constant of the line and
linear ($\propto \omega$) dispersion.  As the fractional width of the third stop band is much larger than the first, the pump can be placed at a propagating frequency while $3 \omega_p$ is blocked.
%

\noindent\textbf{Figure \ref{fig:circuit}  Circuit for paramp gain and noise measurements.}
The pump tone is
produced by a low phase noise synthesizer ($< 150$~dBc at 10~MHz offset), amplified
to a suitable level and filtered using a copper cavity mode bandpass filter.  After a splitter,
the pump tone is attenuated at 4K and filtered using a commercial
combline filter with a bandwidth of 200~MHz around 11.56~GHz.
This bandpass filter is used for the noise measurements,
and provides greater than 70~dB of attenuation for noise on the pump line
at the signal frequencies of interest.
The other output of the
splitter is phase and amplitude adjusted and used to null the pump tone that passes
through the paramp and would otherwise saturate the 4~K HEMT amplifier.  The signal tone
is generated either with another synthesizer or a vector network analyzer.  Its level is reduced by
warm and cold attenuators and coupled to the paramp input using a 20~dB directional coupler
on the base temperature stage of the dilution refrigerator.  A cryogenic isolator is used after the
paramp to absorb noise radiated toward the paramp by the HEMT post-amplification stage.
The noise of the paramp and HEMT is measured with the help of a cryogenic
switch after the isolator that can connect the HEMT amplifier to either
the paramp or 50~\ohm loads at the base temperature and 4~K stages.
The loss of the isolator was measured to be 0.8~dB.
The switch has negligible loss.
After further amplification at room temperature, the signal is measured using
either a spectrum analyzer or
the network analyzer.  The diagram does not
include coax line losses. All coaxes below 4~K are superconducting.

\noindent\textbf{Figure \ref{fig:gain}  Measured and calculated gain}.
(Top) This plot shows a gain profile calculated using the coupled mode equations for FWM
including the dispersion of the device determined using a transmission line model of the periodic loading structure.  A parameter quantifying the nonlinearity of the line and the frequency shift of the dispersion curve due to the cross phase modulation from the pump
tone were adjusted for a reasonable match to the data.
(Second and third) these plots
show the measured ratio of pump on over pump off transmission (gray lines) for two pump powers
(-8.0 and -9.4~dBm at the input of the paramp).  The blue lines are the measured data smoothed by averaging over 60~MHz.  The large peak at 11.9~GHz is an artifact arising from the shift to lower frequency of the transmission dip produced by the periodic loading.
(bottom) The spacing between the gain ripples, shown on an expanded scale, corresponds
to the electrical length of the NbTiN transmission line.

\noindent\textbf{Figure \ref{fig:SNR}  Noise compared with the HEMT amplifier }.
(Left) Increase in signal to
noise ratio of a weak microwave tone applied to the input of the paramp.  With the pump off (green)
the noise floor is limited by the HEMT amplifier.  With the pump on (blue) the signal gain is 18.6~dB and
signal noise ratio has increased by 7.8~dB.  The red lines are a fit to the spectrum analyzer's
response to a monochromatic signal.
The pump on data were taken with $f_s =\,$9.3672~GHz, while the pump off data used $f_s =\,$9.3845~GHz.
(Right) Noise referred to the input of the paramp (blue) and the HEMT amplifier (green) in
photon units versus the calibrator temperature.  The circuit configuration for this measurement is shown in the inset.  The 50~$\Omega$ resistor is mounted on a temperature controlled stage connected to the mixing chamber and terminates a 0.85~mm diameter NbTi coax line.  The noise signal is added to the pump at the input of the paramp using a 3~dB hybrid coupler.
The total loss between the resistor and the paramp input was $4.1 \pm 0.5$~dB, where we have taken the insertion loss of hybrid to lie between the room temperature measured value of 1~dB and zero.
An isolator is used between the noise calibrator and the hybrid to avoid heating from reflected pump power.

\newpage

\newpage

\newpage

\newpage

\end{document}


\setlength{\baselineskip}{24pt}

\newcommand{\jpl}{${}^*$}
\newcommand{\cit}{${}^\dagger$}
\newcommand{\titlefont}[1]{{\Large{\textbf{\textsf{#1}}}}}
\newcommand{\auth}[1]{{\textbf{\textsf{#1}}}}
\newcommand{\addr}[1]{{\small \textit{#1}}}

\begin{center}
\titlefont{Supplementary Information for:  ``A Wideband, Low-Noise Superconducting Amplifier with High Dynamic Range''}\\
\vspace{0.1in}
\auth{Byeong Ho Eom}\cit,
\auth{Peter K. Day}\jpl,
\auth{Henry G. LeDuc}\jpl,
\auth{Jonas Zmuidzinas}\cit\jpl,
\\
\vspace{0.1in}
{\small (Dated: \today)}
\end{center}

\vspace{0.1in}

\noindent
{\cit}\addr{California Institute of Technology,
Pasadena CA 91125, USA}

\noindent
{\jpl}\addr{Jet Propulsion Laboratory, California Institute of Technology, Pasadena CA 91109, USA}

\vspace{0.2in}

\bigskip

\noindent\textbf{Resonator Measurements of the nonlinear response of TiN and NbTiN films}

\bigskip

The dissipation of a system can be sensitively measured using resonant techniques.
Here we present measurements of
microresonators made from thin TiN and NbTiN films, driven with
input powers large enough to cause the resonances to exhibit nonlinear behavior due to the
current dependence of the kinetic inductance.  Our aim is to examine the relationship
between the nonlinear inductance and dissipation in these films.

\begin{figure}
\begin{center}
\includegraphics[width=4in]{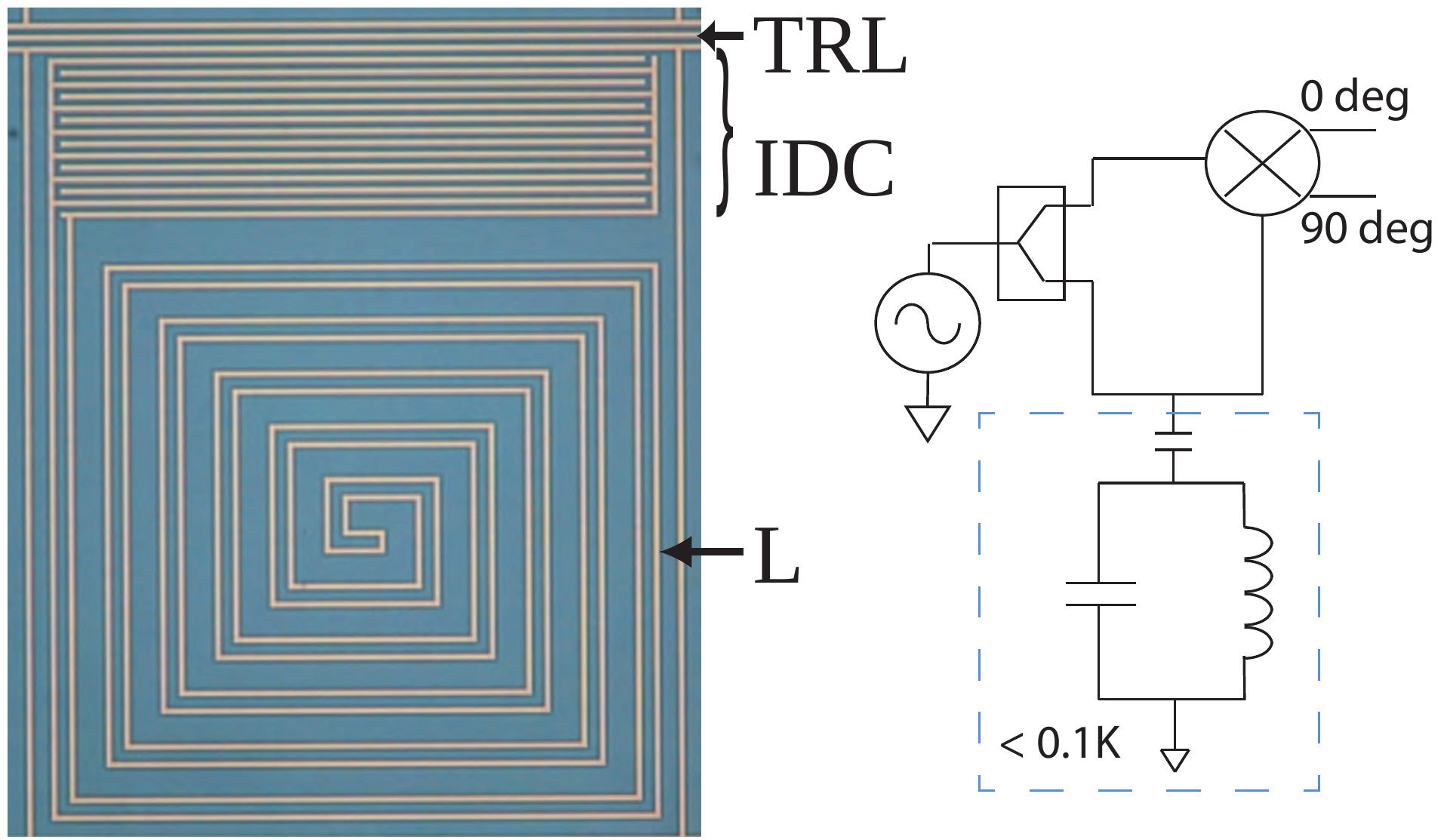}
\includegraphics[width=2in]{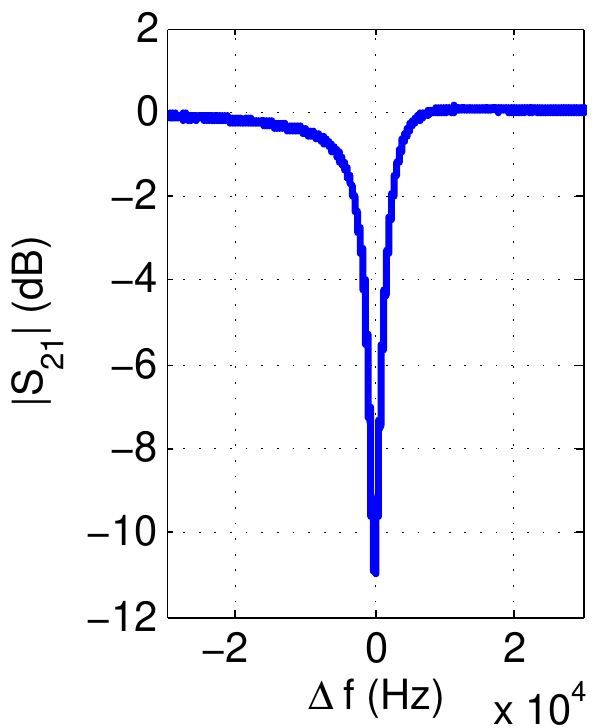}
\caption{  \textbf{ Microresonator geometry and circuit.  (left)}  The
lumped element resonator and microwave feedline (TRL) were patterned from single TiN and NbTiN films
on high resistivity silicon substrates.
The capacitor (IDC) of the resonator consists of
planar interdigital electrodes, and the inductor (L) is a 5 \micron
wide, 50 nm thick trace arranged in a (magnetically) noninductive spiral. The transition
temperatures of the TiN and NbTiN
films were 4K and 14K respectively.
\textbf{(center)}  A homodyne detector circuit measures the complex transmission of the resonator.
The equivalent circuit of the resonator is
a capacitively coupled tank circuit that shunts the readout transmission line, producing a dip in
the transmission at the resonance frequency of 2.2~GHz \textbf{(right)}.
}\label{fig:SIspiralres}
\end{center}
\vspace{-.7cm}
\end{figure}

The measurements were performed on resonators with a ``lumped element'' geometry described
in \SIfig{fig:SIspiralres}, which were weakly coupled capacitively to a finite width ground
plane CPW feedline.
The quality factor $Q_c$ describing the strength of the feedline coupling was $6 \times 10^5$, and the internal quality factors describing resonator losses, $Q_i$ were $5 \times
10^6$ and $2.5 \times 10^6$ for the TiN and NbTiN resonators, respectively.
The devices were cooled in a dilution refrigerator to below 100 mK.

The resonance curves were measured with a homodyne detection circuit (\SIfig{fig:SIspiralres}).
The driving signal was produced by a synthesizer and an I/Q demoduator
was used to measure the complex transmission coefficient.  Use of this
configuration rather than a vector network analyzer allowed the excitation frequency to be swept in both
the upward and downward directions by supplying a low--frequency
($<$ 1 Hz) triangle--wave voltage to the frequency modulation input
of the synthesizer.  It has been found in studies\cite{Yurke_2006} of
similar nonlinear superconducting resonators that
at large enough drive power the system undergoes a bifurcation beyond which
the resonance curves exhibit discontinuous jumps and become hysteretic.  This
behavior occurs whenever the nonlinear dissipation is sufficiently low\cite{Yurke_2006}.
Beyond the bifurcation power, the resonance is more thoroughly explored by downward
frequency sweeps, as these track the resonance
frequency as it is shifted downward by the nonlinearity.  For this reason,
we concentrate here on the downward sweeping direction.

\begin{figure}
\begin{center}
\includegraphics[width=6in]{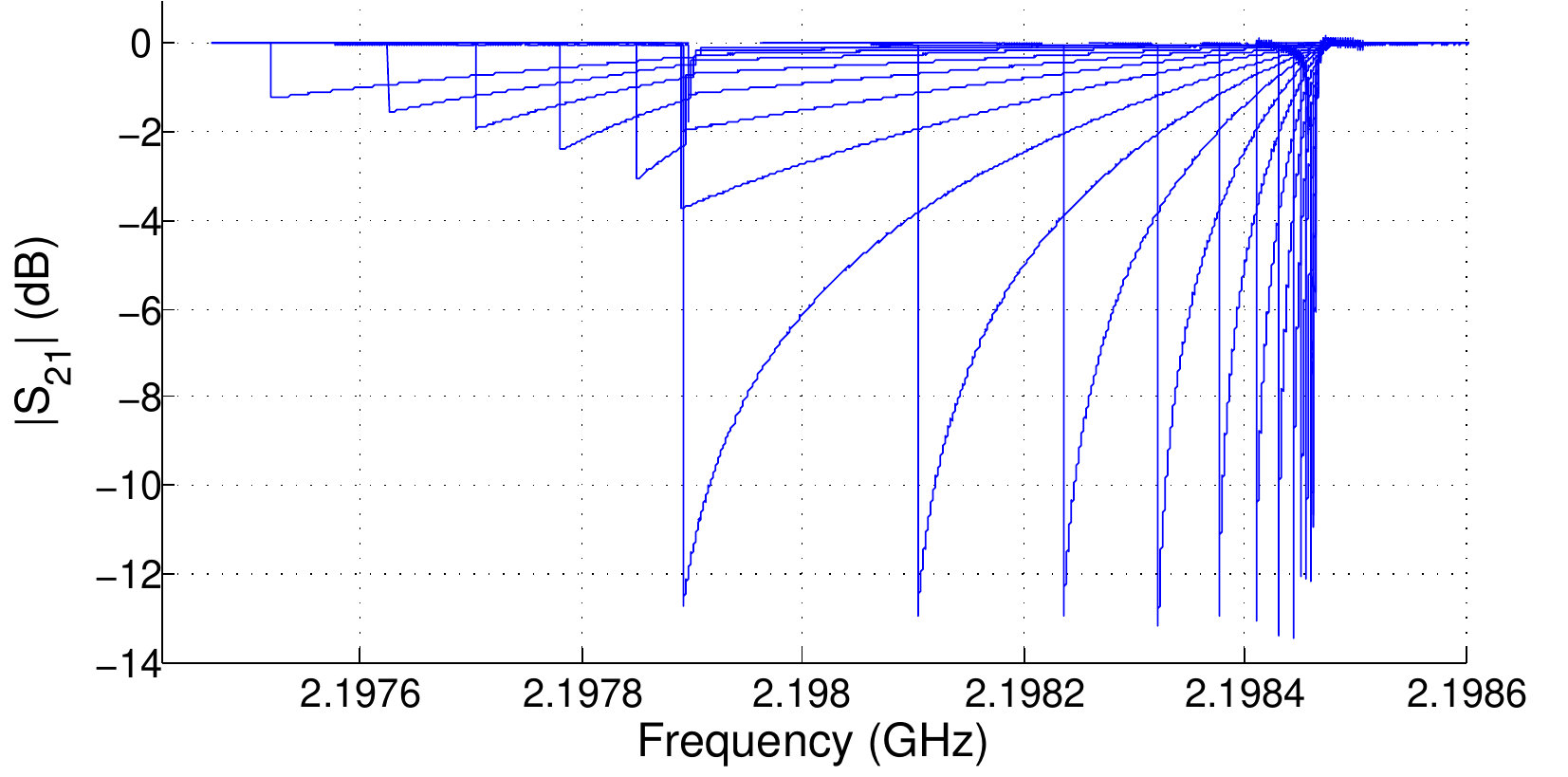}
\caption{  This figure illustrates
the effect of increasing microwave readout power
on the resonance line shape of the 2.2~GHz lumped-element TiN resonator.
As the power is increased, the microwave current in the resonator causes
the inductance to increase, shifting the frequency downwards;
this is classic "soft-spring" Duffing oscillator behavior.
Interestingly, as indicated by the constant depth of the resonance,
dissipation of the resonator changes very little over a wide range of drive
power -- in fact, the change
in resistance of the TiN is $< 2 \times 10^{-4}$ smaller than the change in
its reactance in this range.  Feedline powers are -46 to
-84 dBm in steps of -2 dBm.  The current in the resonator is $\sim 1\,$mA
at the onset of dissipation.} \label{fig:rescurves}
\end{center}
\vspace{-.7cm}
\end{figure}

The data in \SIfig{fig:rescurves} show the highly distorted resonance curves that
result when the resonator is driven well beyond the bifurcation.  The behavior can be
modeled using a current dependent shift
of the resonance frequency $\delta f_r(I)$ and internal quality factor
$Q_i(I)$\cite{Yurke_2006}, where $I$
is the microwave current circulating in the resonator.  In terms of these functions,
the feedline transmission is
\be
S_{21}(f, I) = 1 - \frac{ Q_t(I) / Q_c}{1 + 2iQ_t(I) [f_0 + \delta f_r(I) - f] /
f_0}, \label{eqn:res}
\ee
where $Q_r^{-1}(I) = Q_i^{-1}(I) + Q_c^{-1}$.
The internal current can be expressed in terms of the feedline drive power $P_f$ and $S_{21}$
as
\be
I^2 = 2 Q_c \frac{ |1 - S_{21}|^2}{Z_0} P_f,
\label{eqn:I}
\ee
where $Z_0$ is the impedance of the resonator.  The implicit set of
equations \ref{eqn:res} and \ref{eqn:I} can be solved for $S_{21}$.  For a downward frequency sweep,
at a particular feedline power,
the minimum of $|S_{21}|$ corresponds
to the shifted resonance frequency $f_r(I) = f_0 + \delta f_r(I)$.
At these points $S_{21}$ is real and is equal to $1 - Q_r(I)/Q_c$.
For upward sweeps past the bifurcation power, $S_{21}$ never becomes real, as
the system jumps discontinuously over the resonance condition.

Over a range of drive power the resonance dips shown in \SIfig{fig:rescurves} have
approximately constant depth, hence the resonator dissipation remains unchanged in this range.
In fact, no additional dissipation is observed up to a drive power which is approximately 100 times the bifurcation power.  On further
increasing the feedline power, the resonance dip does
abruptly become much shallower, indicating a sudden onset of dissipation.
Estimation of the resonator microwave current at that point gives $I \approx 1$~mA.

\begin{figure}
\begin{center}
\includegraphics[width=6in]{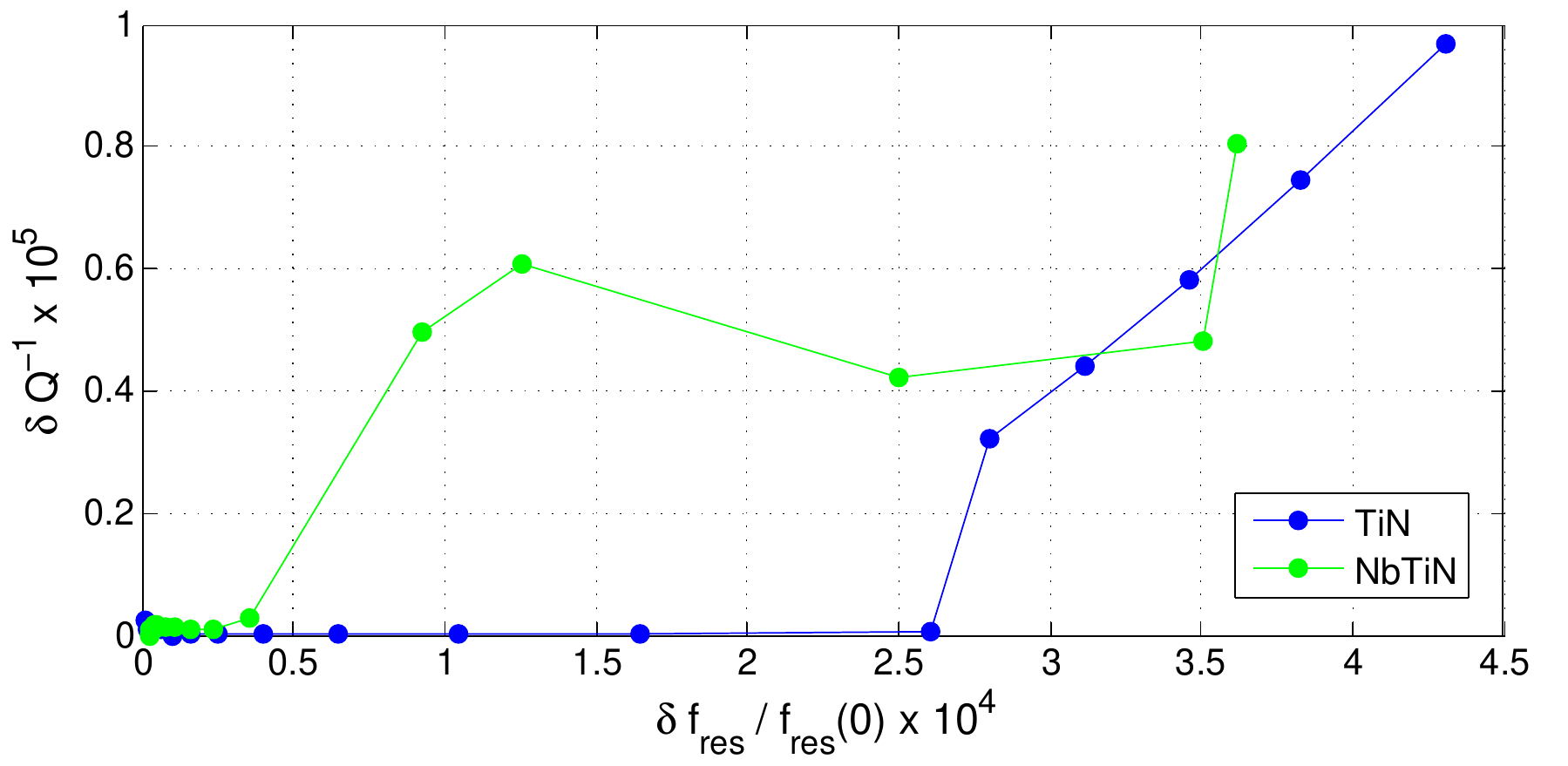}
\caption{ Frequency shift versus change in loss of TiN and NbTiN resonators at the shifted
resonance frequencies for various values of the feedline power.  For TiN, $Q^{-1}$ remains
unchanged to the measurement precision until $\delta f_{res}/f_{res}(0) = 2.6 \times 10^{-4}$,
which corresponds to a internal circulating current of $\approx 1$~mA.  Beyond that point the
loss increases roughly linearly with the frequency shift with a slope $\delta Q^{-1} /
(\delta f/f)$ consistent with an increase in the quasiparticle density in the film.  The NbTiN
resonator shows a reduced range with no dissipation increase and a more complicated behavior
at larger nonlinearity.  For both resonators the loss remains less than $10^{-5}$ over the
range studied.}\label{fig:highpdfdq}
\end{center}
\vspace{-.7cm}
\end{figure}

Figure \SIfig{fig:highpdfdq} shows the change in the resonator loss $\delta Q^{-1}$
versus resonance frequency shift
for the data of \SIfig{fig:rescurves} and for similar measurements on an NbTiN resonator.
For both materials there is a range of drive power over which
there is no increase the dissipation, to within the measurement accuracy, while
significant resonance frequency shifts are observed.
This behavior is in stark contrast to the NbN films used by Abdo~\etal~\cite{Abdo_2006}
and other studies\cite{Dahm_1997,Golosovsky_1995,Chin_1992}
which show strong increases in dissipation as the power is increased.

The magnitude of the nonlinearity reached before dissipation is observed is
$\xi = \delta f_{res}/f_{res} = 2.6 \times 10^{-4}$ for the TiN resonator.  If we wish to
operate the parametric amplifier below this limit, the value of $\xi$ would translate into a
requirement on the transmission line length.  Expressing the gain using the
approximation $f_s \approx f_p$ and assuming perfect phase match $\Delta \beta = -2 \Delta \theta / L$,
we expect $G_s = \exp{( 2 \xi \theta)}/4$, where $\theta = 2\pi L / \lambda_g$ is the phase length in radians of the transmission line with physical length $L$ and guide wavelength $\lambda_g$\cite{Hansryd02}.
For 20~dB gain, we would need $\theta \approx 3 \xi^{-1}$, a readily realizable length for frequencies
in the GHz range, especially
given that the phase velocity on the transmission line may be $\le 0.1c$.  Note that the amplifier
may also be operated at higher nonlinearity, as long as the increase in nonlinear dissipation
remains small, as it does over the entire range of the data in the figure.
In addition to the loss of the superconductor, the tones propagating on the transmission line will be attenuated by losses including
dielectric loss, which is dominated
by two-level systems\cite{Gao08a} at very low temperature.  This loss will be completely
negligible because $\xi Q_i \gg 1$, as can be seen by noting that $Q_i > 10^6$ for resonators made from
TiN\cite{Leduc_2010} or NbTiN\cite{Barends08b}.

\bigskip

\noindent\textbf{Coupled mode equations for predicting the gain of the paramp}

\bigskip

The wave equation for the current $I$ in the transmission line is
\be
\frac{\partial^2 I}{\partial z^2} - \frac{\partial}{\partial t} \left[ \mathcal{L}(I) \mathcal{C}\,
\frac{\partial I}{\partial t}\right] = 0\; ,
\ee
where $\mathcal{L}(I)$ and $\mathcal{C}$ are the inductance and capacitance per unit length, and the
current dependence of the inductance is given by
\be
\mathcal{L}(I) = \mathcal{L}_0 \left( 1 + \frac{I^2}{I^2_\ast} \right).
\ee
As in the standard treatment of waves interacting in nonlinear optical media, we express
the total current in terms of a number of frequency components,
\be
I = \frac{1}{2} \left( \sum_n A_n(z) e^{i(k_n z - \omega_n t)} + {\rm c.c.} \right),
\label{eqn:Isum}
\ee
where the slowly varying complex mode amplitudes $A_n$ satisfy
\be
\left| \frac{d^2 A_n}{dz^2} \right| \ll \left| k_n \frac{d A_n}{dz} \right|.
\ee
The $I^2 dI/dt$ nonlinearity connects combinations of four frequencies.
Hence a general discussion of parametric amplification in
a Kerr medium includes four frequencies in the sum in eqn.~\ref{eqn:Isum}: two
pump tones at $\omega_{p1}$ and $\omega_{p2}$, a weak signal at $\omega_s$ and a
generated idler at $\omega_i = \omega_{p1} + \omega_{p2} - \omega_s$.  Here we
specialize to the case of degenerate four-wave mixing, $\omega_{p1} = \omega_{p2}$.
Evolution of the mode amplitudes $A_p$, $A_s$ and $A_i$ inside the transmission line
is then governed by three coupled mode equations\cite{Agrawalbook},
\begin{eqnarray}
\frac{ dA_p}{dz} & = & \frac{ i k_p}{I'^2_\ast} \left[ \left( |A_p|^2 + 2|A_s|^2 + 2|A_i|^2 \right) A_p
+ 2 A_s A_i A_p^\ast e^{i \Delta \beta z} \right], \nonumber \\
\frac{ dA_s}{dz} & = & \frac{ i k_s}{I'^2_\ast} \left[ \left( |A_s|^2 + 2|A_i|^2 + 2|A_p|^2 \right) A_s
+ A_i^\ast A_p^2 e^{-i \Delta \beta z} \right], \nonumber \\
\frac{ dA_i}{dz} & = & \frac{ i k_i}{I'^2_\ast} \left[ \left( |A_i|^2 + 2|A_s|^2 + 2|A_p|^2 \right) A_i
+ A_s^\ast A_p^2 e^{-i \Delta \beta z} \right],
\label{eqn:coupledmode}
\end{eqnarray}
where the low-power propagation mismatch $\Delta\beta = k_s + k_i - 2k_p$, and $I'^2_\ast = 2 I^2_\ast /
\alpha$.

The number of equations in \ref{eqn:coupledmode} could be increased if we were to include
mixing processes involving, for example, the pump harmonics $(2n+1)\omega_p$, $n = 1,\, 2\, \ldots$, and
combinations such as $2n\omega_p + \omega_s$, etc.  We assume here that these higher frequency tones
either fall in a stop band of the stepped impedance structure and do not propagate, or at least that
the dispersion engineering of the line creates enough phase mismatch at those frequencies that they
do not interact coherently with the lower frequency modes.

For $\omega_s \approx \omega_p$ and using the undepleted pump approximation, $d|A_p|/dz = 0$,
eqns.~\ref{eqn:coupledmode} yield analytical results for the signal power gain $G_s$ and idler
conversion efficiency $G_i$, which may be expressed as\cite{Stolen_1982,Hansryd02}
\begin{eqnarray}
G_s & = & \frac{|A_s(L)|^2}{|A_s(0)|^2} = 1 + \left[ \frac{ k_p |A_p|^2}{I'^2_\ast\, g} \sinh{g L} \right]^2 \\
G_i & = & G_s - 1,
\end{eqnarray}
with parametric gain coefficient $g = -\Delta \beta( \Delta \beta / 4 + k_p |A_p|^2 / I'^2_\ast)$.  In
the case of no dispersion, $\Delta\beta = 0$, $g = 0$ implies that
\be
G_s = 1 + \left( \frac{k_p |A_p|^2}{ I'^2_\ast} \right) = 1 + (\Delta \theta)^2,
\ee
where $\Delta \theta$ is the phase shift of the pump tone.  The gain in this case is quadratic in the
length of the transmission line, rather than exponential, owing to a phase mismatch that results from
the nonlinearity (as can be seen from the factors of 2 in eqn.~\ref{eqn:coupledmode}, the phase shifts of
the signal and idler tones are twice that of the pump tone.)  If $\Delta \beta = -k_p |A_p|^2 / I'^2_\ast$,
the linear phase mismatch compensates that due to the nonlinearity and the exponential gain regime is
accessed with
\be
G_s \approx \frac{1}{4} \exp{(2 \Delta\theta)}.
\ee

\begin{figure}
\begin{center}
\includegraphics[width=3in]{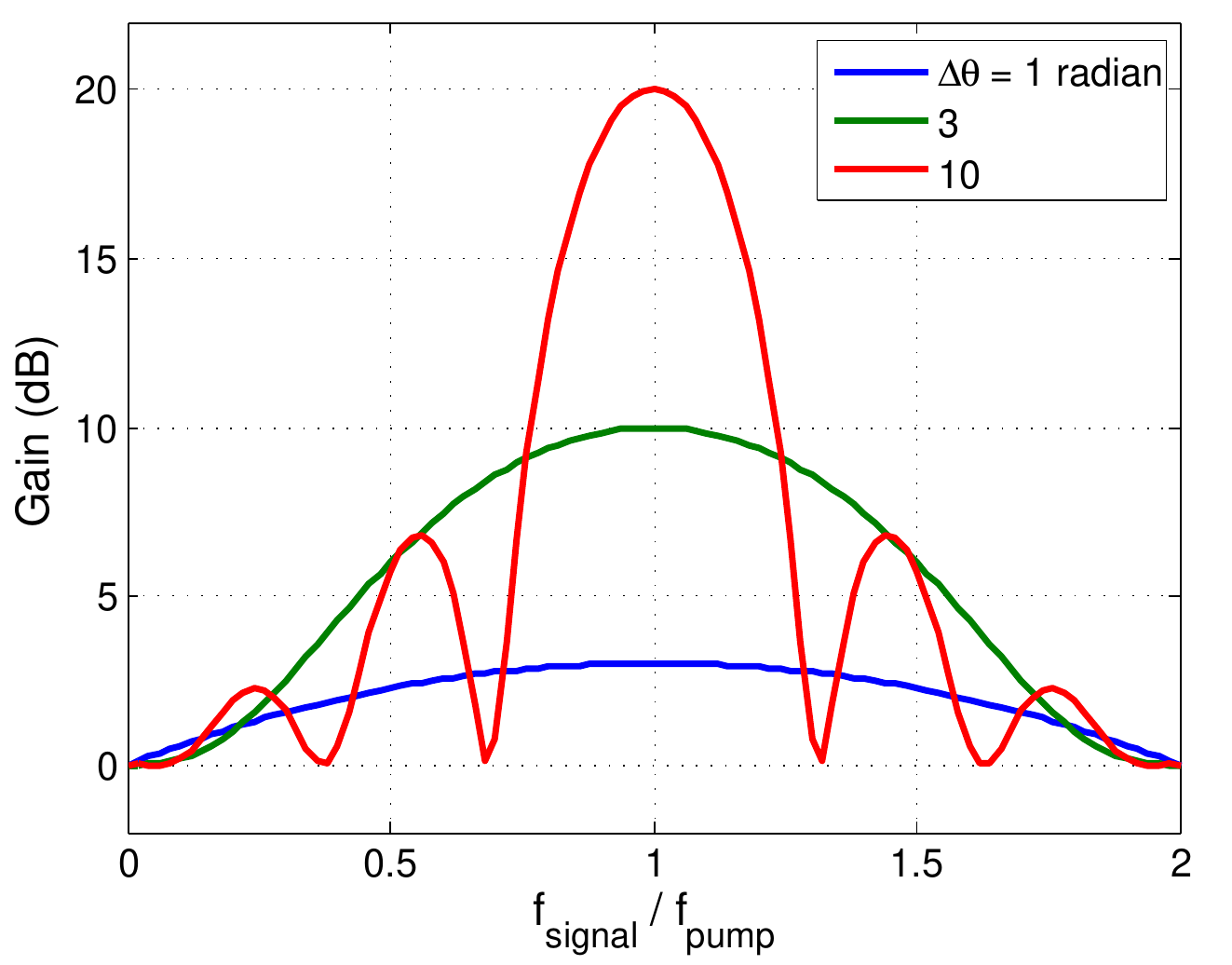}
\includegraphics[width=3in]{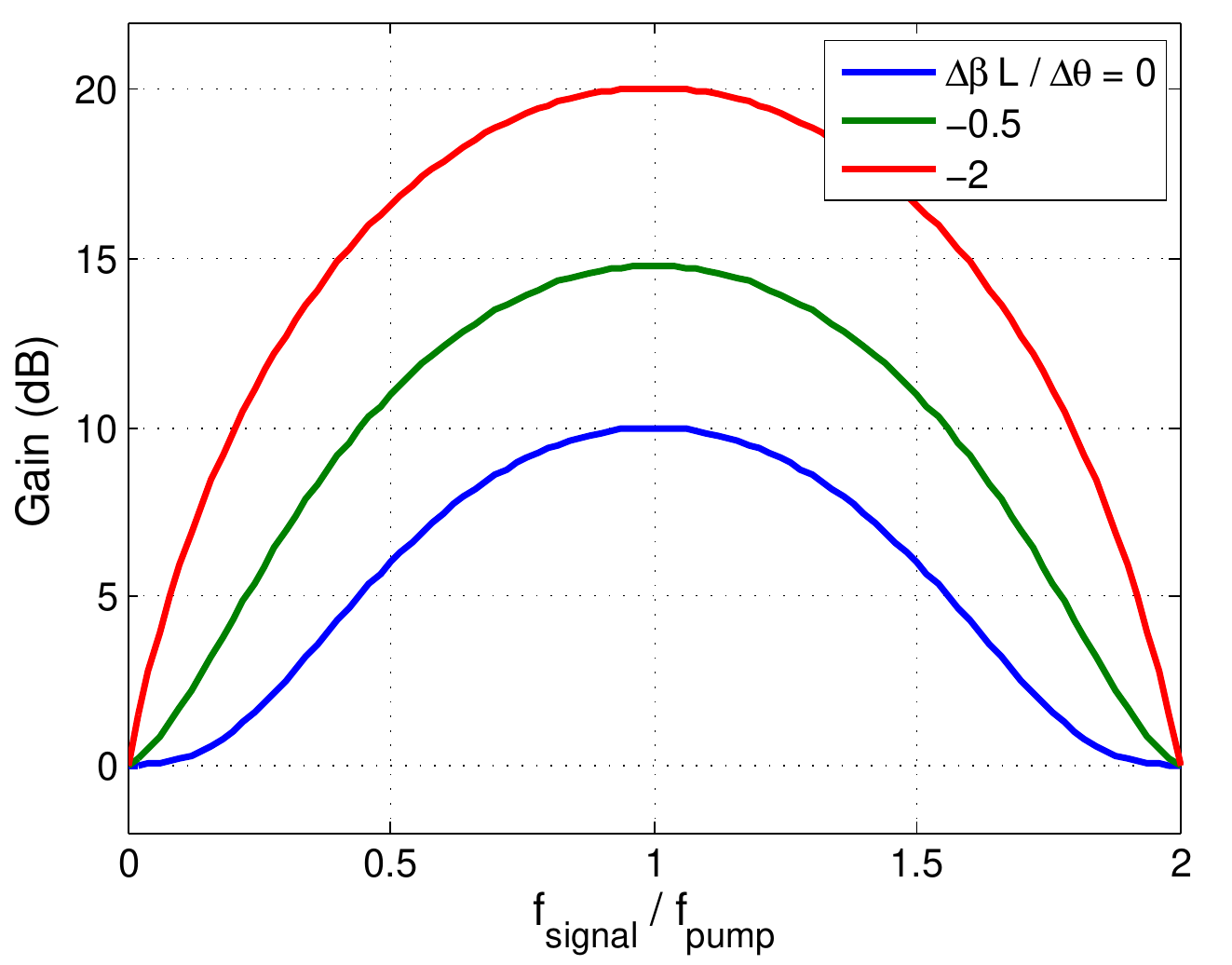}
\caption{ Calculation of the paramp gain using the coupled mode equations \ref{eqn:coupledmode} for
\textbf{(left:)} no dispersion and various values of the nonlinear phase shift $\Delta \theta$; and
\textbf{(right:)} fixed $\Delta \theta = 3$~radians and various values of $\Delta\beta$.}\label{fig:coupledmode}
\end{center}
\vspace{-.7cm}
\end{figure}

The kinetic inductance paramp described in this paper may operate well outside the frequency
range where $\omega_s \approx \omega_p$ is valid, so to predict the gain and bandwidth we integrate eqns.~\ref{eqn:coupledmode} numerically.  The results of that calculation for several idealized cases are shown
in \SIfig{fig:coupledmode}.  Here we assume that the frequencies other than those of the pump, signal and idler
are blocked by engineered stop bands or large dispersion at those frequencies.  For simplicity, we
ignore the narrow stop band near the pump that would be associated with introducing dispersion at that
frequency.  We also neglect the detailed form of the dispersion near $\omega_p$, taking $\Delta\beta$ to be either a negative constant or zero.  The calculation results uphold the very broadband nature of the amplifier.  In
the case of no dispersion, the 3-dB bandwidth narrows, but remains above 25\% for 20~dB peak gain.  For
optimal $\Delta\beta$, 20~dB peak gain may be realized with a phase shift of only 3 radians, and the 3-dB
bandwidth is greater than an octave.

\pagebreak

\noindent\textbf{Device fabrication}

\bigskip

The NbTiN itself is deposited by dc reactive magnetron- sputtering from a 3~inch diameter NbTi metal target.  The films are sputtered using constant current (.55-.8A) in a background of Argon and Nitrogen.  The process pressure is maintained at 5 mTorr using a downstream conductance controller and the flow rates are set using mass flow controllers.  The deposition system is a load-locked UHV system with a base pressure $\sim 5 \times 10^{-10}$ Torr. The TiN films are deposited following the procedures discussed in \cite{Leduc_2010}.

\printbibliography